\documentstyle[12pt]{article}
\input epsf

\catcode`\@=11
\newcommand{\smalllineskip}{\baselineskip=12pt}
\newcommand{\fcaption}[1]{
        \refstepcounter{figure}
        \setbox\@tempboxa = \hbox{\small Fig.~\thefigure. #1}
        \ifdim \wd\@tempboxa > 5in
           {\begin{center}
        \parbox{5in}{\small\smalllineskip Fig.~\thefigure. #1}
            \end{center}}
        \else
             {\begin{center}
             {\small Fig.~\thefigure. #1}
              \end{center}}
        \fi}

\def\spose#1{\hbox to 0pt{#1\hss}}
\def\lsim{\mathrel{\spose{\lower 3pt\hbox{$\mathchar"218$}}
 \raise 2.0pt\hbox{$\mathchar"13C$}}}
\def\gsim{\mathrel{\spose{\lower 3pt\hbox{$\mathchar"218$}}
 \raise 2.0pt\hbox{$\mathchar"13E$}}}

\global\arraycolsep=2pt
\begin{document}
\topskip 2cm
\begin{titlepage}
\begin{flushright}
CERN-TH/96-120\\
hep-ph/9605256
\end{flushright}
\vspace{0.5cm}

\begin{center}
{\large\bf A Fresh Look at the B Semileptonic Branching Ratio\\
and Beauty Lifetimes}\\
\vspace{2.5cm}
{\large Matthias Neubert}\\
\vspace{0.5cm}
{\sl Theory Division, CERN, CH-1211 Geneva 23, Switzerland}\\
\vspace{2.5cm}
\vfil
\begin{abstract}
We discuss two problems in the theory of heavy-quark decays: an
understanding of the semileptonic branching ratio of $B$ mesons, and
of the lifetime ratio $\tau(\Lambda_b)/\tau(B^0)$. We also present a
model-independent study of spectator contributions to the lifetimes
of beauty hadrons.
\end{abstract}
\end{center}

\vspace{1.0cm}
\begin{center}
\it To appear in the Proceedings of\\
Les Rencontres de Physique de la Vall\'ee d'Aoste\\
``Results and Perspectives in Particle Physics''\\
La Thuile, Aosta Valley, Italy, March 1996
\end{center}
\vspace{2.0cm}
\noindent
CERN-TH/96-120\\
May 1996
\end{titlepage}

\section{Introduction}

Inclusive decays of heavy hadrons have two advantages from the
theoretical point of view: first, bound-state effects related to the
initial state can be accounted for in a systematic way using the
heavy-quark expansion \cite{Chay}--\cite{MaWe}; secondly, the fact
that the final state consists of a sum over many hadronic channels
eliminates bound-state effects related to the properties of
individual hadrons. This last feature is based on the hypothesis of
quark--hadron duality, i.e.\ the assumption that cross sections and
decay rates are calculable in QCD after an ``averaging'' procedure
has been applied \cite{PQW}. This assumption has been tested
experimentally using data on hadronic $\tau$ decays \cite{Maria}. The
theory of inclusive decays of heavy hadrons proved to be very
successful (for a recent review, see Ref.~\cite{MNnew}). For
instance, it explains {\em a posteriori\/} the success of the parton
model in describing inclusive semileptonic decays of heavy hadrons.
However, we shall address here two potential problems of this theory:
the semileptonic branching ratio of $B$ mesons, and the short
lifetime of the $\Lambda_b$ baryon.

The inclusive decay width of a hadron $H_b$ containing a $b$ quark
can be written as the forward matrix element of the imaginary part of
the transition operator, $m_{H_b}\,\Gamma(H_b\to X) = \mbox{Im}\,
\langle H_b|\,{\bf T}\,|H_b\rangle$, where ${\bf T}$ is given by
\begin{equation}\label{Top}
   {\bf T} = i\!\int{\rm d}^4 x\,T\{\,
   {\cal L}_{\rm eff}(x),{\cal L}_{\rm eff}(0)\,\} \,.
\end{equation}
For the case of semileptonic and non-leptonic decays, the effective
weak Lagrangian is
\begin{eqnarray}
   {\cal L}_{\rm eff} &=& - {4 G_F\over\sqrt{2}}\,V_{cb}\,
    \bigg\{ c_1(m_b)\,\Big[
    \bar d'_{\rm L}\gamma_\mu u_{\rm L}\,
    \bar c_{\rm L}\gamma^\mu b_{\rm L} +
    \bar s'_{\rm L}\gamma_\mu c_{\rm L}\,
    \bar c_{\rm L}\gamma^\mu b_{\rm L} \Big] \nonumber\\
   &&\phantom{ - {4 G_F\over\sqrt{2}}\,V_{cb}\, }
    \mbox{}+ c_2(m_b)\,\Big[
    \bar c_{\rm L}\gamma_\mu u_{\rm L}\,
    \bar d'_{\rm L}\gamma^\mu b_{\rm L} +
    \bar c_{\rm L}\gamma_\mu c_{\rm L}\,
    \bar s'_{\rm L}\gamma^\mu b_{\rm L} \Big] \nonumber\\
   &&\phantom{ - {4 G_F\over\sqrt{2}}\,V_{cb}\, }
    \mbox{}+ \sum_{\ell=e,\mu,\tau}
    \bar\ell_{\rm L}\gamma_\mu\nu_\ell\,
    \bar c_{\rm L}\gamma^\mu b_{\rm L} \bigg\} + \mbox{h.c.} \,,
\end{eqnarray}
where $q_{\rm L}=\frac{1}{2}(1-\gamma_5) q$ denotes a left-handed
quark field, $d'$ and $s'$ are the Cabibbo-rotated down- and
strange-quark fields, and we have neglected $b\to u$ transitions. The
Wilson coefficients $c_1$ and $c_2$ take into account the QCD
corrections arising from the fact that the effective Lagrangian is
written at a renormalization scale $\mu=m_b$ rather than $m_W$. The
combinations $c_\pm=c_1\pm c_2$ are renormalized multiplicatively.

\begin{figure}[htb]
   \epsfxsize=8cm
   \centerline{\epsffile{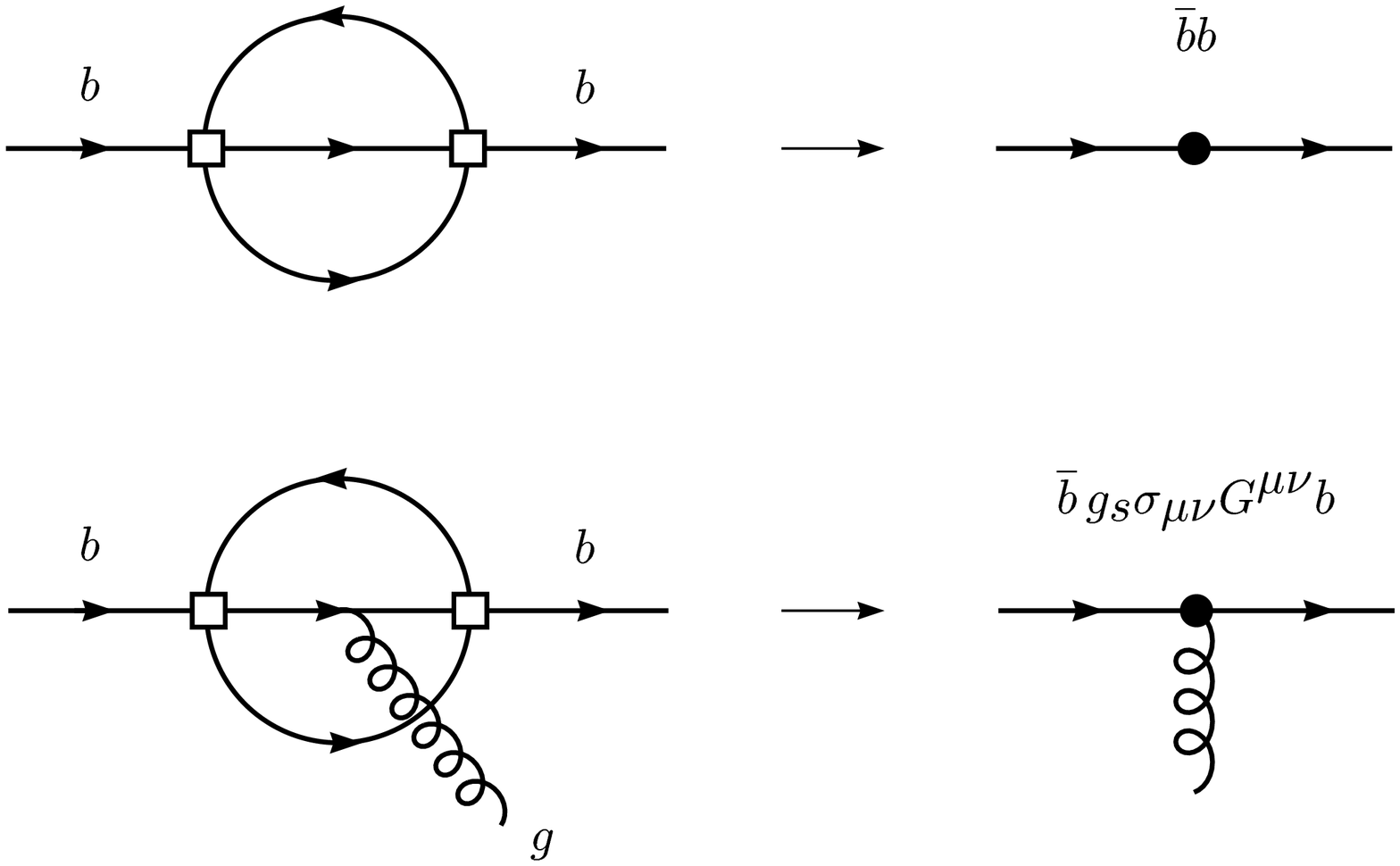}}
   \centerline{\parbox{14cm}{\fcaption{\label{fig:Toper}
Perturbative contributions to the transition operator ${\bf T}$
(left), and the corresponding operators in the OPE (right). The open
squares represent a four-fermion interaction of the effective
Lagrangian ${\cal L}_{\rm eff}$, while the black dots represent
local operators in the heavy-quark expansion.
   }}}
\end{figure}

In perturbation theory, some contributions to the transition operator
are given by the two-loop diagrams shown on the left-hand side of
Fig.~\ref{fig:Toper}. Since the energy release in the decay of a $b$
quark is large, it is possible to construct an Operator Product
Expansion (OPE) for the bilocal transition operator (\ref{Top}), in
which it is expanded as a series of local operators with increasing
dimension, whose coefficients contain inverse powers of the $b$-quark
mass. The operator with the lowest dimension is $\bar b b$. It arises
from integrating over the internal lines in the first diagram. There
is no independent operator with dimension four, since the only
candidate, $\bar b\,i\rlap{\,/}D\,b$, can be reduced to $\bar b b$ by
using the equations of motion. The first new operator is $\bar
b\,g_s\sigma_{\mu\nu} G^{\mu\nu} b$ and has dimension five. It arises
from diagrams in which a gluon is emitted from one of the internal
lines, such as the second diagram in Fig.~\ref{fig:Toper}.

In the next step, the forward matrix elements of the local operators
in the OPE are systematically expanded in inverse powers of the
$b$-quark mass, using the heavy-quark effective theory (HQET)
\cite{review}. The purpose of doing this expansion is to introduce a
minimal set of hadronic parameters which are independent of $m_b$.
The result is that any inclusive decay rate of a hadron $H_b$ can be
written in the form \cite{Bigi}--\cite{MaWe}
\begin{equation}\label{generic}
   \Gamma(H_b\to X_f) = {G_F^2 m_b^5\over 192\pi^3}\,
   \Bigg\{ c_3^f\,\bigg( 1 - {\mu_\pi^2(H_b)\over 2 m_b^2} \bigg)
   + c_5^f\,{\mu_G^2(H_b)\over 2 m_b^2}
   + O(1/m_b^3) \Bigg\} \,,
\end{equation}
where $\mu_\pi^2(H_b)$ and $\mu_G^2(H_b)$ parametrize the matrix
elements of the kinetic-energy and the chromo-magnetic operators,
respectively, and $c_n^f$ are calculable coefficient functions (which
also contain the relevant CKM matrix elements) depending on the
quantum numbers $f$ of the final state. For semileptonic and
non-leptonic decays, the coefficients $c_3^f$ have been calculated at
one-loop order \cite{Hoki,BSLnew1}, and the coefficients $c_5^f$ at
tree level \cite{Bigi,FLNN}. The relevant combinations of the
hadronic parameters $\mu_\pi^2(H_b)$ and $\mu_G^2(H_b)$ for $B$
mesons and $\Lambda_b$ baryons can be determined from the spectrum of
heavy-hadron states \cite{MNnew,liferef}.

\section{Semileptonic Branching Ratio}
\label{sec:Bsl}

The semileptonic branching ratio of $B$ mesons is defined as
\begin{equation}
   B_{\rm SL} = {\Gamma(B\to X\,e\,\bar\nu)\over
   \sum_\ell \Gamma(B\to X\,\ell\,\bar\nu) + \Gamma_{\rm NL}
   + \Gamma_{\rm rare}} \,,
\end{equation}
where $\Gamma_{\rm NL}$ and $\Gamma_{\rm rare}$ are the inclusive
rates for non-leptonic and rare decays, respectively. Measurements of
this quantity have been performed by various experimental groups. The
status of the results is controversial, as there is a discrepancy
between low-energy measurements performed at the $\Upsilon(4s)$
resonance and high-energy measurements performed at the $Z^0$
resonance. The average value at low energies is $B_{\rm SL}=(10.37\pm
0.30)\%$ \cite{Tomasz}, whereas high-energy measurements give $B_{\rm
SL}^{(b)}=(11.11\pm 0.23)\%$ \cite{Pascal}. The superscript $(b)$
indicates that this value refers not to $B$ mesons, but to a mixture
of $b$ hadrons. Correcting for this fact, we find the slightly larger
value $B_{\rm SL}=(11.30\pm 0.26)\%$ \cite{MNnew}. The discrepancy
between the low- and high-energy measurements of the semileptonic
branching ratio is therefore larger than three standard deviations.
If we take the average and inflate the error to account for this
fact, we obtain $B_{\rm SL}=(10.90\pm 0.46)\%$. An important aspect
in understanding this result is charm counting, i.e.\ the measurement
of the average number $n_c$ of charm hadrons produced per $B$ decay.
Recently, two new (preliminary) measurements of this quantity have
been performed. The CLEO Collaboration has presented the value
$n_c=1.16\pm 0.05$ \cite{Tomasz}, and the ALEPH Collaboration has
reported the result $n_c=1.20\pm 0.08$ \cite{ALEPHnc}. The average is
$n_c=1.17\pm 0.04$.

In the parton model, $B_{\rm SL}\simeq 13\%$ and $n_c\simeq 1.15$
\cite{Alta}. Whereas $n_c$ is in agreement with experiment, the
semileptonic branching ratio is predicted to be too large. With the
establishment of the $1/m_Q$ expansion the non-perturbative
corrections to the parton model could be computed, and they turned
out to be too small to improve the prediction \cite{baff}. The
situation has changed recently, when it was found that higher-order
perturbative corrections lower the value of $B_{\rm SL}$
significantly \cite{BSLnew1}. The exact order-$\alpha_s$ corrections
to the non-leptonic width have been computed for $m_c\ne 0$, and an
analysis of the renormalization scale and scheme dependence has been
performed. In particular, it turns out that radiative corrections
increase the partial width $\Gamma(B\to X_{c\bar c s})$ by a large
amount. This has two effects: it lowers the semileptonic branching
ratio, but at the price of a higher value of $n_c$.

\begin{figure}[htb]
\vspace{1.8cm}
   \centerline{
   \epsfysize=5cm\epsffile{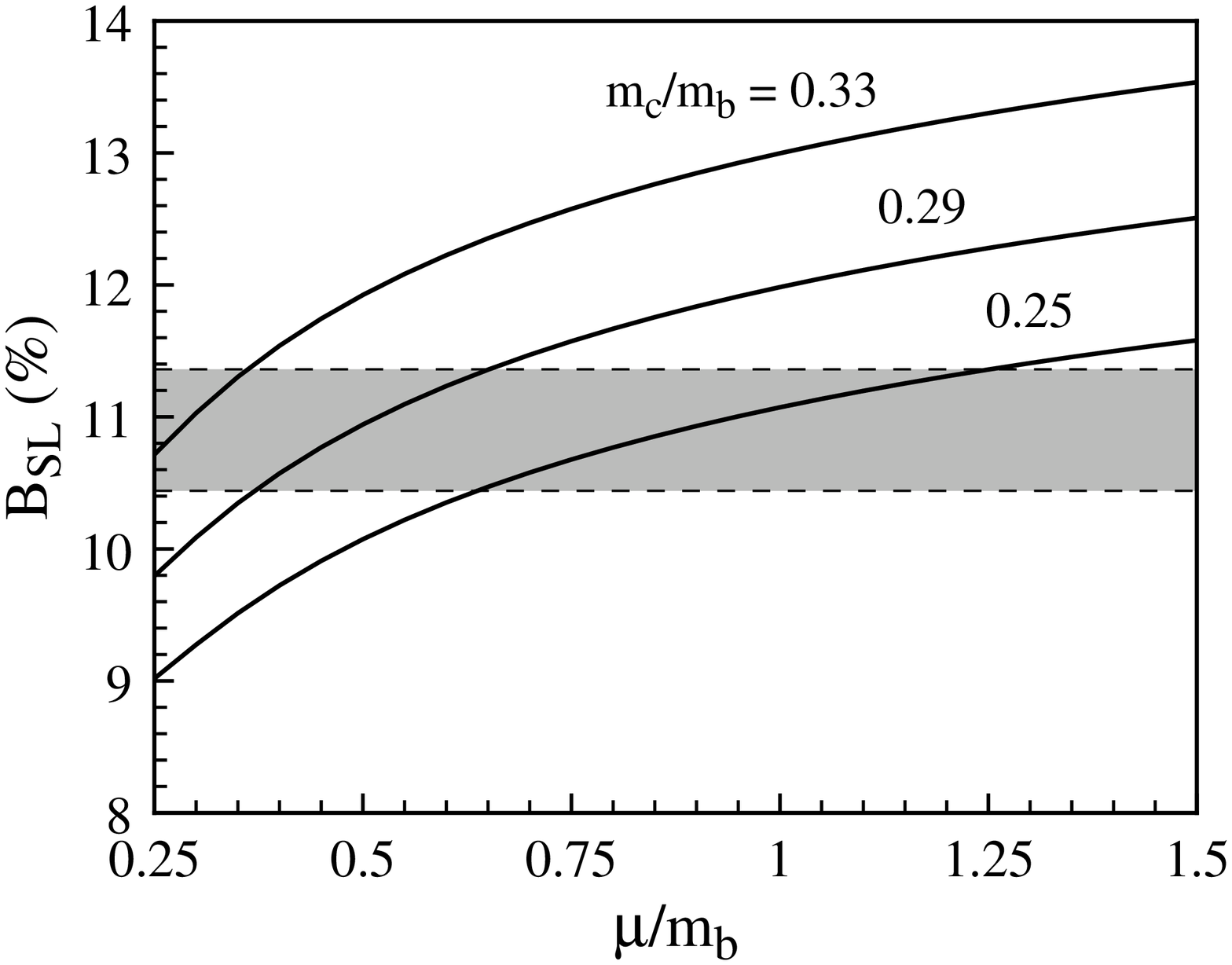}
   \epsfysize=5cm\epsffile{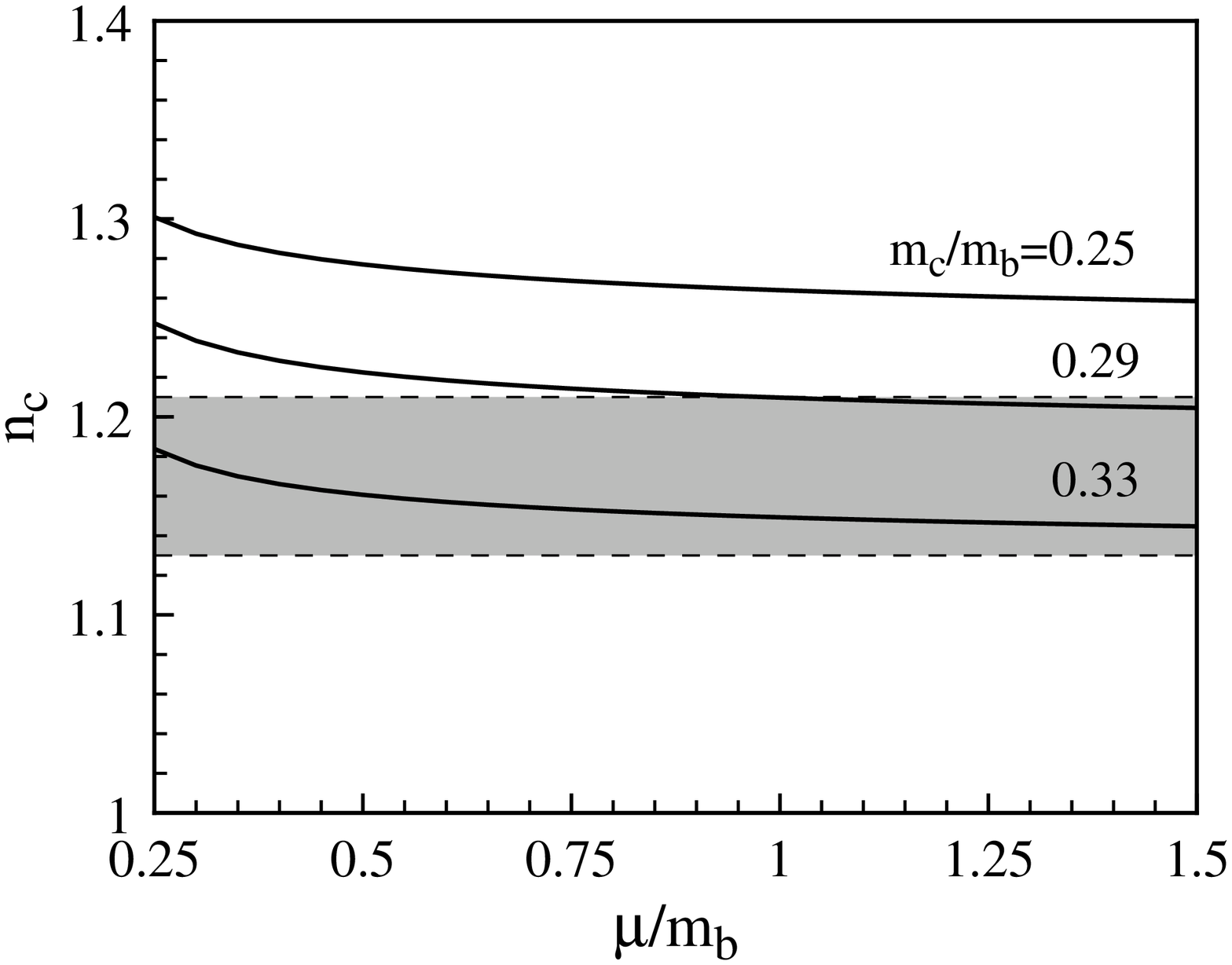}}
   \centerline{\parbox{14cm}{\fcaption{\label{fig:mudep}
Scale dependence of the theoretical predictions for the semileptonic
branching ratio and $n_c$. The bands show the average experimental
values.
   }}}
\vspace{-1.8cm}
\end{figure}

\begin{figure}[htb]
   \epsfxsize=7cm
   \centerline{\epsffile{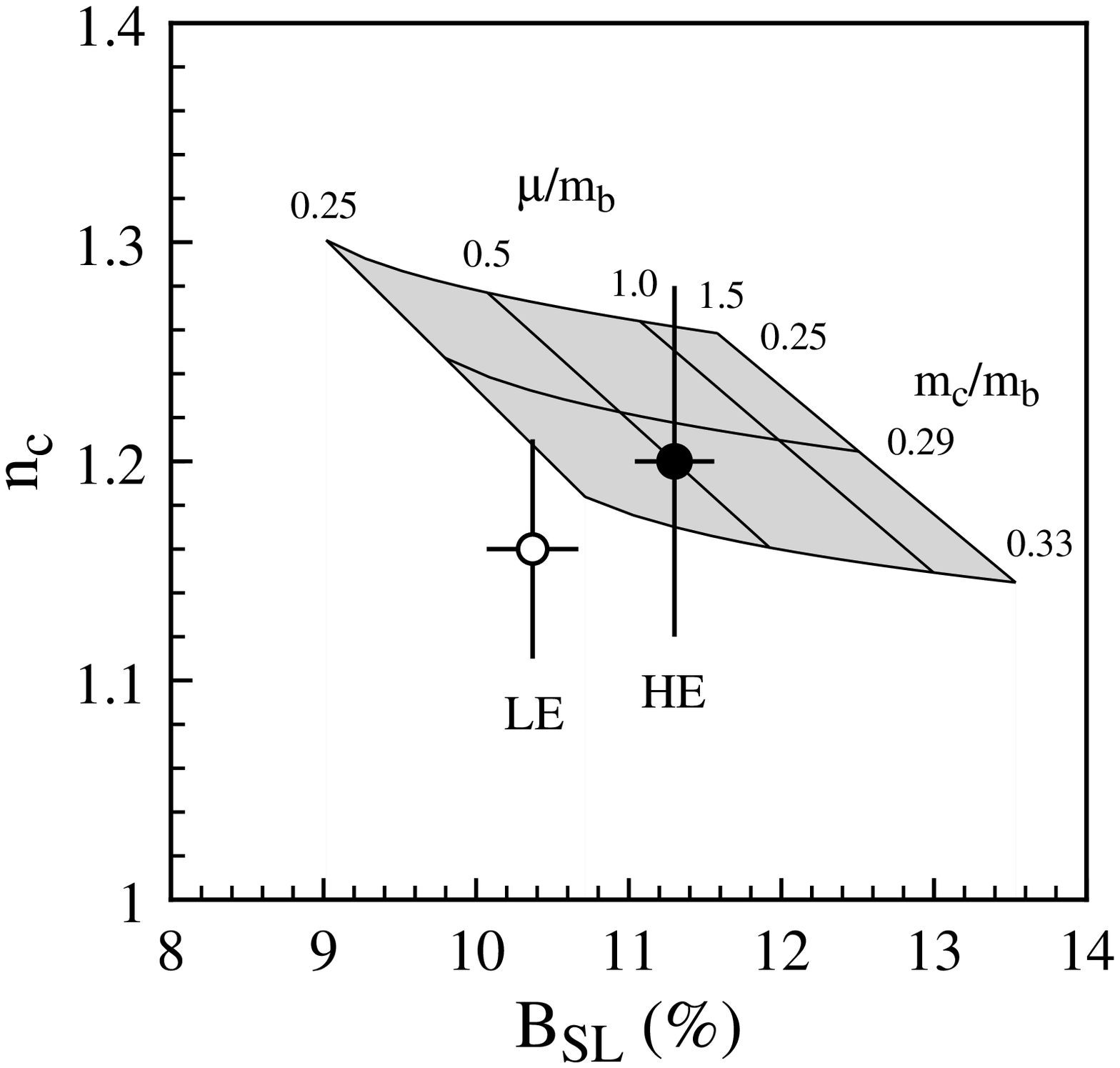}}
   \centerline{\parbox{14cm}{\fcaption{\label{fig:BSL}
Combined theoretical predictions for the semileptonic branching ratio
and charm counting as a function of the quark-mass ratio $m_c/m_b$
and the renormalization scale $\mu$. The data points show the average
experimental values for $B_{\rm SL}$ and $n_c$ obtained in low-energy
(LE) and high-energy (HE) measurements, as discussed in the text.
   }}}
\end{figure}

The original analysis of Bagan et al.\ has recently been corrected in
an erratum \cite{BSLnew1}. Here we shall present the results of an
independent numerical analysis using the same theoretical input (for
a detailed discussion, see Ref.~\cite{Chris}). The semileptonic
branching ratio and $n_c$ depend on the quark-mass ratio $m_c/m_b$
and on the ratio $\mu/m_b$, where $\mu$ is the scale used to
renormalize the coupling constant $\alpha_s(\mu)$ and the Wilson
coefficients $c_\pm(\mu)$ appearing in the non-leptonic decay rate.
Below we shall consider several choices for the renormalization
scale. We allow the (one-loop) pole masses of the heavy quarks to
vary in the range $m_b=(4.8\pm 0.2)$~GeV and $m_b-m_c=(3.40\pm
0.06)$~GeV, corresponding to $0.25<m_c/m_b<0.33$. Non-perturbative
effects appearing at order $1/m_b^2$ in the heavy-quark expansion are
described by the single parameter $\mu_G^2(B)$; the dependence on the
parameter $\mu_\pi^2(B)$ is the same for all inclusive decay rates
and cancels out in $B_{\rm SL}$ and $n_c$. For the two choices
$\mu=m_b$ and $\mu=m_b/2$, we obtain
\begin{eqnarray}
   B_{\rm SL} &=& \cases{
    12.0\pm 1.0 \% ;& $\mu=m_b$, \cr
    10.9\pm 0.9 \% ;& $\mu=m_b/2$, \cr} \nonumber\\
   \phantom{ \bigg[ }
   n_c &=& \cases{
    1.21\mp 0.06 ;& $\mu=m_b$, \cr
    1.22\mp 0.06 ;& $\mu=m_b/2$. \cr}
\end{eqnarray}
The uncertainties in the two quantities, which result from the
variation of $m_c/m_b$ in the range given above, are anticorrelated.
Notice that the semileptonic branching ratio has a stronger scale
dependence than $n_c$. This is illustrated in Fig.~\ref{fig:mudep},
where we show the two quantities as a function of $\mu$. By choosing
a low renormalization scale, values $B_{\rm SL}<12\%$ can easily be
accommodated. The experimental data prefer a scale $\mu/m_b\sim 0.5$,
which is indeed not unnatural; it has been estimated that $\mu\gsim
0.32 m_b$ is an appropriate scale to use \cite{LSW}. The combined
theoretical predictions for the semileptonic branching ratio and
charm counting are shown in Fig.~\ref{fig:BSL}. They are compared
with the experimental results obtained from low- and high-energy
measurements. It was argued that the combination of a low
semileptonic branching ratio and a low value of $n_c$ would
constitute a potential problem for the Standard Model
\cite{baff,Buch}. However, with the new experimental and theoretical
numbers, it is only for the low-energy measurements that a small
discrepancy remains between theory and experiment.

\section{Lifetime Ratios of Beauty Hadrons}

The heavy-quark expansion predicts that the lifetimes of all beauty
hadrons agree up to non-perturbative corrections suppressed by at
least two powers of $1/m_b$. By explicit evaluation of the general
result (\ref{generic}) for semi- and non-leptonic decays, one finds
\cite{Chris}
\begin{equation}\label{taucrude}
   {\tau(B^-)\over\tau(B^0)} = 1 + O(1/m_b^3) \,, \qquad
   {\tau(\Lambda_b)\over\tau(B^0)} = 0.98 + O(1/m_b^3) \,.
\end{equation}
These theoretical predictions may be compared with the average
experimental values for the lifetime ratios \cite{Joe}:
$\tau(B^-)/\tau(B^0)=1.02\pm 0.04$ and
$\tau(\Lambda_b)/\tau(B^0)=0.78\pm 0.05$. Whereas the lifetime ratio
of charged and neutral $B$ mesons is in good agreement with the
theoretical prediction, the low value of the lifetime of the
$\Lambda_b$ baryon is surprising.

\begin{figure}[htb]
   \epsfxsize=8cm
   \centerline{\epsffile{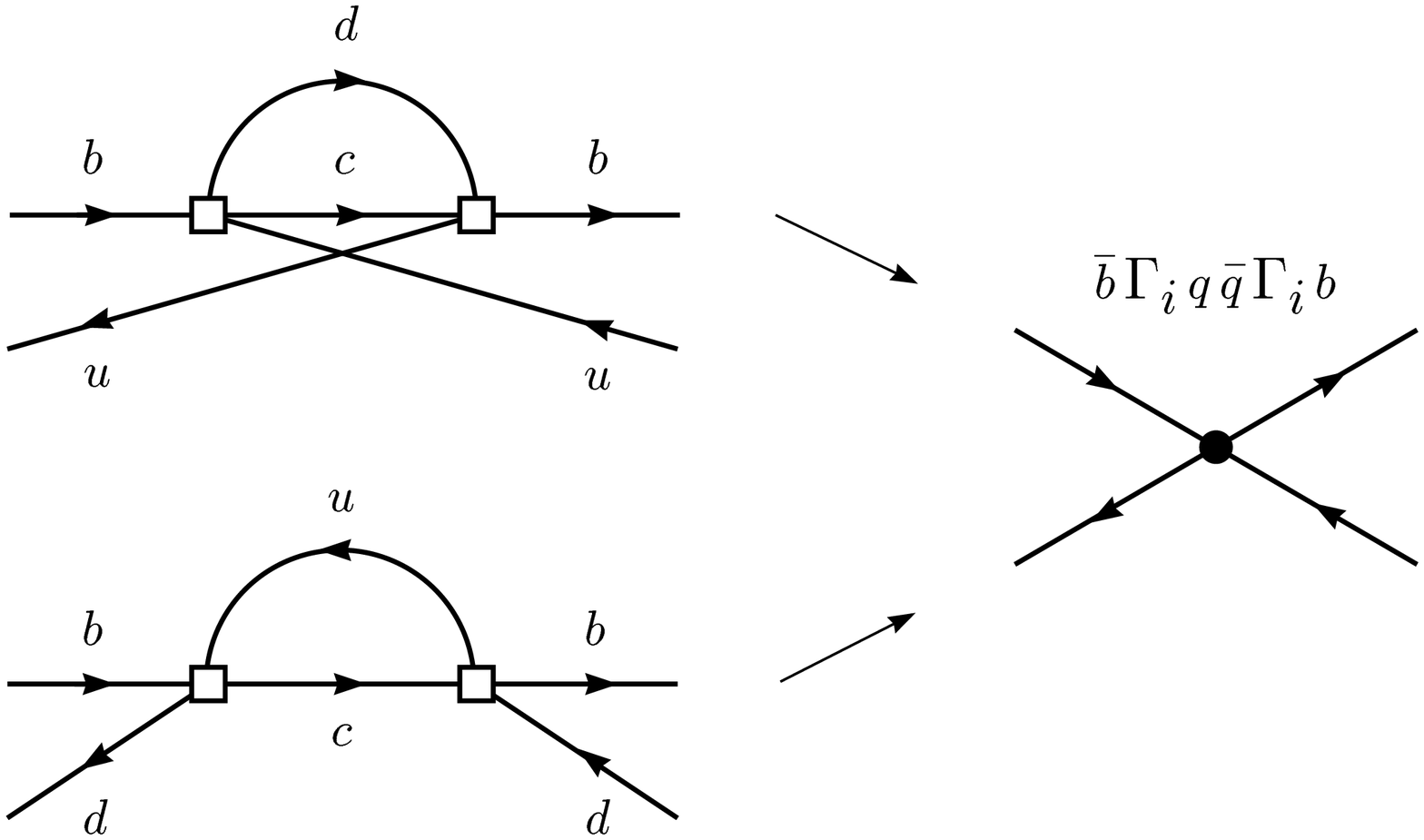}}
   \centerline{\parbox{14cm}{\fcaption{\label{fig:Tspec}
Spectator contributions to the transition operator ${\bf T}$ (left),
and the corresponding operators in the OPE (right). Here $\Gamma_i$
denotes some combination of Dirac and colour matrices.
   }}}
\end{figure}

To understand the structure of the lifetime differences requires to
go further in the $1/m_b$ expansion. ``Spectator effects''
\cite{Gube}, i.e.\ contributions from decays in which a light
constituent quark participates in the weak process, contribute
directly to the differences in the decay widths of different beauty
hadrons. They are suppressed because the $b$ quark and a light quark
in the heavy hadron need to be close together. Since the portion of
the volume that the $b$ quark occupies inside the hadron is of order
$(\Lambda_{\rm QCD}/m_b)^3$, such effects appear only at third order
in the heavy-quark expansion, and it might seem safe to neglect them
altogether. However, as a result of the difference in the phase space
for $2\to 2$-body reactions as compared to $1\to 3$-body decays,
spectator effects are enhanced by a factor of order $16\pi^2$. This
can be seen from Fig.~\ref{fig:Tspec}, which shows that the
corresponding contributions to the transition operator ${\bf T}$
arise from one-loop rather than two-loop diagrams. The situation is
different from gluonic dimension-six operators of the type $\bar
b\gamma_\mu(i D_\nu G^{\mu\nu}) b$, which appear at the same order in
the heavy-quark expansion. Such operators arise from decays in which
the light spectators interact only softly. Since their matrix
elements are flavour blind and not enhanced by phase space, they can
be safely neglected.

If one cuts the internal lines in Fig.~\ref{fig:Tspec}, one obtains
the spectator contributions to the decay operator
${\bf\Gamma}=2\,\mbox{Im}\,{\bf T}$. The corresponding spectator
effects are referred to as Pauli interference and $W$ exchange
\cite{Gube}. The spectator contributions to the non-leptonic width of
beauty mesons and baryons are given by the matrix elements of the
local operator \cite{Chris}
\begin{eqnarray}\label{Gspec}
   {\bf\Gamma}_{\rm spec}
   &=& {2 G_F^2 m_b^2\over 3\pi}\,|V_{cb}|^2\,(1-z)^2\,\Big\{
    (2 c_+^2 - c_-^2)\,O_{\rm V-A}^u
    + 3 (c_+^2 + c_-^2)\,T_{\rm V-A}^u \Big\} \nonumber \\
   &-& {2 G_F^2 m_b^2\over 9\pi}\,|V_{cb}|^2\,(1-z)^2\,\bigg\{
    (2 c_+ - c_-)^2\,\Big[ (1 + \textstyle\frac{z}{2})\,
    O_{\rm V-A}^{d'} - (1+2z)\,O_{\rm S-P}^{d'} \Big] \nonumber\\
   &&\phantom{ {2 G_F^2 m_b^2\over\pi}\,|V_{cb}|^2\,(1-z)^2\, }
    \mbox{}+ \textstyle\frac{3}{2} (c_+ + c_-)^2\,\Big[
    (1 + \frac{z}{2})\,T_{\rm V-A}^{d'}
    - (1+2z)\,T_{\rm S-P}^{d'} \Big] \bigg\} \nonumber\\
   &-& {2 G_F^2 m_b^2\over 9\pi}\,|V_{cb}|^2\,\sqrt{1-4 z}\,\bigg\{
    (2 c_+ - c_-)^2\,\Big[ (1-z)\,O_{\rm V-A}^{s'}
    - (1+2z)\,O_{\rm S-P}^{s'} \Big] \nonumber\\
   &&\phantom{ {2 G_F^2 m_b^2\over\pi}\,|V_{cb}|^2\,(1-z)^2\, }
    \mbox{}+ \textstyle\frac{3}{2} (c_+ + c_-)^2\,\Big[
    (1-z)\,T_{\rm V-A}^{s'} - (1+2z)\,T_{\rm S-P}^{s'} \Big]
    \bigg\} \,,
\end{eqnarray}
where $z=m_c^2/m_b^2$. The local four-quark operators appearing in
this expression are defined by
\begin{eqnarray}\label{4qops}
   O_{\rm V-A}^q &=& \bar b_{\rm L}\gamma_\mu q_{\rm L}\,
    \bar q_{\rm L}\gamma^\mu b_{\rm L}\,, \qquad\quad\,\,
   O_{\rm S-P}^q = \bar b_{\rm R}\,q_{\rm L}\,
    \bar q_{\rm L}\,b_{\rm R} \,, \nonumber\\
   T_{\rm V-A}^q &=& \bar b_{\rm L}\gamma_\mu t_a q_{\rm L}\,
    \bar q_{\rm L}\gamma^\mu t_a b_{\rm L} \,, \qquad
   T_{\rm S-P}^q = \bar b_{\rm R}\,t_a q_{\rm L}\,
    \bar q_{\rm L}\,t_a b_{\rm R} \,,
\end{eqnarray}
where $t_a$ are the generators of colour SU(3). These operators are
renormalized at the scale $m_b$, which will be implicit in our
discussion below. We note that in the limit $z=0$ our result agrees
with Ref.~\cite{liferef}, and with the corresponding calculations for
charm decays \cite{Gube}.

The hadronic matrix elements of the four-quark operators in
(\ref{4qops}) contain the non-perturbative physics of the spectator
contributions to inclusive decays of beauty hadrons. In most previous
analyses of spectator effects, these matrix elements have been
estimated using simplifying assumptions. For the matrix elements
between $B$-meson states the vacuum saturation approximation
\cite{SVZ} was assumed, i.e.\ the matrix elements of the four-quark
operators were evaluated by inserting the vacuum inside the current
products, in which case they are determined by the square of the
decay constant $f_B$ of the $B$ meson. In order to avoid such
model-dependent assumptions, we define without loss of generality
\cite{Chris}:
\begin{eqnarray}\label{Biepsi}
   {1\over 2 m_B}\,\langle B|\,O_{\rm V-A}^q\,|B\rangle
   &=& B_1\,{f_B^2 m_B\over 8} \,, \qquad
   \displaystyle{1\over 2 m_B}\,
   \langle B|\,O_{\rm S-P}^q\,|B\rangle
   = B_2\,{f_B^2 m_B\over 8} \,, \nonumber\\
   {1\over 2 m_B}\,\langle B|\,T_{\rm V-A}^q\,|B\rangle
   &=& \varepsilon_1\,{f_B^2 m_B\over 8} \,, \qquad\,\,
   \displaystyle{1\over 2 m_B}\,
   \langle B|\,T_{\rm S-P}^q\,|B\rangle
   = \varepsilon_2\,{f_B^2 m_B\over 8} \,.
\end{eqnarray}
The values of the dimensionless hadronic parameters $B_i$ and
$\varepsilon_i$ are currently not known; ultimately, they may be
calculated using some field-theoretic approach such as lattice gauge
theory or QCD sum rules. The vacuum saturation approximation
corresponds to setting $B_i=1$ and $\varepsilon_i=0$ (at some scale
$\mu$, where the approximation is believed to be valid\footnote{
Usually, the vacuum saturation approximation is applied at a typical
hadronic scale $\mu_{\rm had}\ll m_b$. The values of $B_i$ and
$\varepsilon_i$ at the scale $m_b$ are then affected by
renormalization effects. Taking, for instance, $\alpha_s(\mu_{\rm
had})=0.5$ (corresponding to $\mu_{\rm had}\sim 0.75$~GeV), we find
$B_1(m_b)=B_2(m_b)\simeq 1.01$ and $\varepsilon_1(m_b)=
\varepsilon_2(m_b)\simeq -0.05$.}). For real QCD, however, it is
known that $B_i=O(1)$ and $\varepsilon_i=O(1/N_c)$, where $N_c$ is
the number of colours.

In the case of $\Lambda_b$ baryons, we find it convenient to
introduce the operators ($i,j$ are colour indices)
$\widetilde O_{\rm V-A} = \bar b_{\rm L}^i\gamma_\mu q_{\rm L}^j\,
\bar q_{\rm L}^j\gamma^\mu b_{\rm L}^i$ and $\widetilde O_{\rm S-P}
 = \bar b_{\rm R}^i\,q_{\rm L}^j\,\bar q_{\rm L}^j\,b_{\rm R}^i$
instead of $T_{\rm V-A}$ and $T_{\rm S-P}$. They are related by the
colour Fierz identity $T=-\frac{1}{6}\,O+\frac{1}{2}\,\widetilde O$.
The heavy-quark spin symmetry, i.e.\ the fact that interactions with
the spin of the heavy quark decouple as the heavy-quark mass tends to
infinity, implies the relations $\langle\Lambda_b|\,O_{\rm S-P}\,
|\Lambda_b\rangle = -\frac{1}{2}\,\langle\Lambda_b|\,O_{\rm V-A}\,
|\Lambda_b\rangle$ and $\langle\Lambda_b|\,\widetilde O_{\rm S-P}\,
|\Lambda_b\rangle = -\frac{1}{2}\,\langle\Lambda_b|\,
\widetilde O_{\rm V-A}\,|\Lambda_b\rangle$ \cite{Chris}. This leaves
us with two independent matrix elements of the operators $O_{\rm
V-A}$ and $\widetilde O_{\rm V-A}$. The analogue of the vacuum
insertion approximation in the case of baryons is the valence-quark
assumption, in which the colour of the quark fields in the operators
is identified with the colour of the quarks inside the baryon. Since
the colour wave-function for a baryon is totally antisymmetric, the
matrix elements of $O_{\rm V-A}$ and $\widetilde O_{\rm V-A}$ differ
in this approximation only by a sign. Hence, we define a parameter
$\widetilde B$ by
\begin{equation}\label{Btildef}
   \langle\Lambda_b|\,\widetilde O_{\rm V-A}\,|\Lambda_b\rangle
   \equiv - \widetilde B\,\langle\Lambda_b|\,O_{\rm V-A}\,
   |\Lambda_b\rangle \,,
\end{equation}
with $\widetilde B=1$ in the valence-quark approximation. For the
baryon matrix element of $O_{\rm V-A}$ itself, our parametrization is
guided by the quark model. We write
\begin{equation}\label{rdef}
   {1\over 2 m_{\Lambda_b}}\,\langle\Lambda_b|\,O_{\rm V-A}\,
   |\Lambda_b\rangle \equiv - {f_B^2 m_B\over 48}\,r \,,
\end{equation}
where in the quark model $r$ is the ratio of the squares of the wave
functions determining the probability to find a light quark at the
location of the $b$ quark inside the $\Lambda_b$ baryon and the $B$
meson, i.e.\ $r=|\psi_{bq}^{\Lambda_b}(0)|^2/|\psi_{b\bar q}^{B}(0)
|^2$ \cite{Gube}. Assuming that the wave functions of the
$\Lambda_b$ and $\Sigma_b$ baryons are the same, the ratio $r$ can be
estimated from the ratio of the spin splittings between $\Sigma_b$
and $\Sigma_b^*$ baryons and $B$ and $B^*$ mesons \cite{Rosn}. This
leads to
\begin{equation}\label{rval}
   r = {4\over 3}\,{m_{\Sigma_b^*}^2 - m_{\Sigma_b}^2\over
                    m_{B^*}^2 - m_B^2} \simeq 0.9\pm 0.1 \,,
\end{equation}
where we have taken the baryon mass-splitting to be $m_{\Sigma_b^*}^2
- m_{\Sigma_b}^2\simeq m_{\Sigma_c^*}^2 - m_{\Sigma_c}^2$.

\subsection{Lifetime ratio $\tau(B^-)/\tau(B^0)$}

Because of isospin symmetry, the lifetimes of the charged and neutral
$B$ mesons are the same at order $1/m_b^2$ in the heavy-quark
expansion, and differences arise only from spectator effects. The
explicit calculation of these effects leads to a contribution to the
decay width given by \cite{Chris}
\begin{equation}\label{DeltaGam}
   \Gamma_{\rm spec}(B) = 16\pi^2\,{f_B^2\,m_B\over m_b^3}\,
   \zeta_B\,\Gamma_0 \,;\quad
   \Gamma_0 = {G_F^2 m_b^5\over 192\pi^3}\,|V_{cb}|^2 \,,
\end{equation}
where $\zeta_{B^-}\simeq -0.4 B_1+6.6\varepsilon_1$ and
$\zeta_{B^0}\simeq -2.2\varepsilon_1+2.4\varepsilon_2$. Note the
factor of $16\pi^2$, which arises from the phase-space enhancement of
spectator effects. Since the parton-model result for the total decay
width is $\Gamma(B)_{\rm tot}\simeq 3.7\,\Gamma_0$, the
characteristic scale of spectator contributions is $(2\pi f_B/m_B)^2
\simeq 5\%$. Thus, it is natural that the lifetimes of different
beauty hadrons differ by a few per cent.

The precise value of the lifetime ratio crucially depends on the size
of the hadronic matrix elements. Taking $f_B=200$~MeV for the decay
constant of the $B$ meson, i.e.\ absorbing the uncertainty in this
parameter into the definition of $B_i$ and $\varepsilon_i$, leads to
\cite{Chris}
\begin{equation}
   {\tau(B^-)\over\tau(B^0)} \simeq 1 + 0.03 B_1 - 0.7\varepsilon_1
   + 0.2\varepsilon_2 \,.
\end{equation}
The most striking feature of this result is that the coefficients of
the colour-octet operators $T_{\rm V-A}$ and $T_{\rm S-P}$ are orders
of magnitude larger than those of the colour-singlet operator $O_{\rm
V-A}$. As a consequence, the vacuum insertion approximation, which
was adopted in Ref.~\cite{liferef} to predict that
$\tau(B^-)/\tau(B^0)$ is larger than unity by an amount of order 5\%,
cannot be trusted. With $\varepsilon_i$ of order $1/N_c$, it is
conceivable that the non-factorizable contributions dominate the
result, and without a detailed calculation of the parameters
$\varepsilon_i$ no reliable prediction can be obtained. Given our
present ignorance about the true values of the hadronic matrix
elements, we must conclude that even the sign of the sum of the
spectator contributions cannot be predicted. A lifetime ratio in the
range $0.8<\tau(B^-)/\tau(B^0)<1.2$ could be easily accommodated by
theory. In view of these considerations, the experimental fact that
the lifetime ratio turns out to be very close to unity is somewhat of
a surprise. It implies a constraint on a certain combination of the
colour-octet matrix elements, which reads
$\varepsilon_1-0.3\varepsilon_2=\mbox{few~\%}$.

\subsection{Lifetime ratio $\tau(\Lambda_b)/\tau(B^0)$}

Understanding the low experimental value of the lifetime ratio
$\tau(\Lambda_b)/\tau(B^0)$ is one of the major problems in
heavy-quark theory. We will now discuss the structure of spectator
contributions to this ratio. It is important that the heavy-quark
symmetry allows us to reduce the number of hadronic parameters
contributing to the decay rate of the $\Lambda_b$ baryon from four to
two, and that these parameters are almost certainly positive (unless
the quark model is completely misleading) and enter the decay rate
with the same sign. Thus, unlike the meson case, the structure of
the spectator contributions to the width of the $\Lambda_b$ baryon is
rather simple, and at least the sign of the effects can be predicted
reliably.

It is useful to distinguish between the two cases where one does or
does not allow spectator contributions to enhance the theoretical
prediction for the semileptonic branching ratio of $B$ mesons. As we
have shown in Section~\ref{sec:Bsl}, the theoretical prediction for
$B_{\rm SL}$, which neglects spectator contributions, is slightly
larger than the central experimental value. If spectator effects
increased the prediction for $B_{\rm SL}$ further, this discrepancy
could become uncomfortably large.

If we do not allow for an increase in the value of the semileptonic
branching ratio, the explanation of the low value of
$\tau(\Lambda_b)/\tau(B^0)$ must reside entirely in a low value of
the $\Lambda_b$ lifetime (rather than a large value of the $B$-meson
lifetime). This can be seen by writing
\begin{equation}
   {\tau(\Lambda_b)\over\tau(B^0)} = \tau(\Lambda_b)\,
   \left( {\tau(B^-)\over\tau(B^0)} \right)^{1/2}\,
   {1\over\big[\tau(B^-)\,\tau(B^0)\big]^{1/2}} \nonumber\\
   = {\tau(\Lambda_b)\over B_{\rm SL}}\,\left(
   {\tau(B^-)\over\tau(B^0)} \right)^{1/2}\,\Gamma_{\rm SL}(B) \,,
\end{equation}
where $B_{\rm SL}$ is the average semileptonic branching ratio of
$B$ mesons, and $\Gamma_{\rm SL}(B)$ is the semileptonic width. In
the last step we have replaced the geometric mean
$[\tau(B^-)\,\tau(B^0)]^{1/2}$ by the average $B$-meson lifetime,
which because of isospin symmetry is correct to order $1/m_b^6$ in
the heavy-quark expansion. Since there are no spectator contributions
to the semileptonic rate $\Gamma_{\rm SL}(B)$, and since we do not
allow an enhancement of the semileptonic branching ratio, in order
to obtain a small value for $\tau(\Lambda_b)/\tau(B^0)$ we can
increase the width of the $\Lambda_b$ baryon and/or decrease (within
the experimental errors) the lifetime ratio $\tau(B^-)/\tau(B^0)$.
Allowing for a downward fluctuation of this ratio by two standard
deviations, i.e.\ $\tau(B^-)/\tau(B^0)>0.94$, and using the estimate
of $1/m_b^2$ corrections in (\ref{taucrude}), we conclude that
\begin{equation}\label{didef}
   {\tau(\Lambda_b)\over\tau(B^0)} > 0.97 \times \bigg( 0.98
   - {\Gamma_{\rm spec}(\Lambda_b)\over\Gamma(\Lambda_b)} \bigg)
   = 0.95 - (d_1 + d_2\widetilde B)\,r \,,
\end{equation}
where $\Gamma_{\rm spec}(\Lambda_b)$ is the spectator contribution to
the width of the $\Lambda_b$ baryon. The values of the coefficients
$d_i$ depend on the scale $\mu$, at which the Wilson coefficients
$c_\pm(\mu)$ are renormalized.\footnote{For $\mu\ne m_b$ we take into
account the evolution of the operators between the scales $\mu$ and
$m_b$, so that the parameters defining the matrix elements of the
four-quark operators are always renormalized at $m_b$.}
For $\mu=m_b$ we find $d_1=0.013$ and $d_2=0.022$, whereas for
$\mu=m_b/2$ we obtain the larger values $d_1=0.018$ and $d_2=0.024$.
If we assume that $r$ and $\widetilde B$ are of order unity, we find
that the spectator contributions yield a reduction of the lifetime of
the $\Lambda_b$ baryon by a few per cent, and that
$\tau(\Lambda_b)/\tau(B^0)>0.9$, in contrast with the experimental
result. If we try to push the theoretical prediction by taking the
large value $\widetilde B=1.5$ (corresponding to a violation of the
valence-quark approximation by 50\%) and choosing a low scale
$\mu=m_b/2$, we have to require that $r>r_{\rm min}$ with $r_{\rm
min}=3.1$, 2.2 and 1.3 for $\tau(\Lambda_b)/\tau(B^0)=0.78$, 0.83 and
0.88 (corresponding to the central experimental value and the
$1\sigma$ and $2\sigma$ fluctuations). Hence, unless we allow for an
upward fluctuation of the experimental result by two standard
deviations, we need a value of $r$ that is significantly larger than
the quark-model prediction in (\ref{rval}). A reliable
field-theoretic calculation of the parameters $r$ and $\widetilde B$
is of great importance to support or rule out such a possibility.

On the other hand, the low experimental value of the semileptonic
branching ratio may find its explanation in a low renormalization
scale (see Figs.~\ref{fig:mudep} and \ref{fig:BSL}), or it may be
caused by the effects of New Physics, such as an enhanced rate for
flavour-changing neutral currents of the type $b\to s g$. Hence, one
may be misled in using the semileptonic branching ratio as a
constraint on the size of spectator contributions. Then there is the
possibility to decrease the value of $\tau(\Lambda_b)/\tau(B^0)$ by
increasing the lifetime of the $B^0$ meson, i.e.\ in (\ref{didef}) we
can allow for spectator contributions to the width of the $B^0$
meson. At first sight, this seems to make it possible (with a
suitable choice of $\varepsilon_1$ and $\varepsilon_2$) to gain a
contribution of about $-0.1$, which would take away much of the
discrepancy between theory and experiment. However, the experimental
result for the lifetime ratio $\tau(B^-)/\tau(B^0)$ imposes the
powerful constraint $\varepsilon_1\simeq 0.3\varepsilon_2$. Using
this to eliminate $\varepsilon_1$ from the result, and allowing the
parameters $B_i$ to take values between 0 and 2, we find
\begin{equation}\label{LamBd}
   {\tau(\Lambda_b)\over\tau(B^0)} \simeq 0.98 \pm 0.02
   + 0.15\varepsilon_2 - (d_1 + d_2\widetilde B)\,r
   > 0.88 - (d_1 + d_2\widetilde B)\,r \,,
\end{equation}
where in the last step we have assumed that $|\varepsilon_2|<0.5$,
which we consider to be a conservative bound. Even in this case, a
significant contribution must still come from the parameters $r$ and
$\widetilde B$.

In view of the above discussion, the short $\Lambda_b$ lifetime
remains a potential problem for the heavy-quark theory. If the
current experimental value persists, there are two possibilities:
either some hadronic matrix elements of four-quark operators are
significantly larger than naive expectations based on large-$N_c$
counting rules and the quark model, or (local) quark--hadron duality,
which is assumed in the calculation of lifetimes, fails in
non-leptonic inclusive decays. In the second case, the explanation of
the puzzle lies beyond the heavy-quark expansion. Let us, therefore,
consider the first possibility and give a numerical example for some
possible scenarios. Assume that $\mu=m_b/2$ is an appropriate scale
to use in the evaluation of the Wilson coefficients, and that
$\widetilde B=1.5$. Then, to obtain $\tau(\Lambda_b)/\tau(B^0)=0.8$
without enhancing the prediction for the semileptonic branching ratio
requires $r\simeq 3$, i.e.\ three times larger than the quark-model
estimate in (\ref{rval}). If, on the other hand, we consider $r=1.5$
as the largest conceivable value, we need $\varepsilon_2\simeq -0.5$,
corresponding to a rather large matrix element of the colour-octet
operator $T_{\rm S-P}$. Such a value of $\varepsilon_2$ leads to an
enhancement of the $B$-meson lifetime, and hence to an enhancement of
the semileptonic branching ratio of $B$ mesons, by $\Delta B_{\rm
SL}\simeq 1\%$ \cite{Chris}. As shown in Figs.~\ref{fig:mudep} and
\ref{fig:BSL}, this is still tolerable provided yet unknown
higher-order corrections confirm the use of a low renormalization
scale. Although in both cases some large parameters are needed, we
find it important to note that until reliable field-theoretic
calculations of the matrix elements of four-quark operators become
available, a conventional explanation of the $\Lambda_b$-lifetime
puzzle cannot be excluded.

\section{Conclusions}

The heavy-quark expansion, supplemented by the assumption of
quark--hadron duality, provides the theoretical framework for a
systematic calculation of inclusive decay rates of hadrons containing
a heavy quark. Whereas this formalism works well for the description
of the total decay rate and the lepton and neutrino spectra in
semileptonic decays, two potential problems related to non-leptonic
decays have become apparent in recent years: the low experimental
value of the semileptonic branching ratio of $B$ mesons, and the low
value of the lifetime ratio $\tau(\Lambda_b)/\tau(B^0)$.

We have shown that the semileptonic branching ratio can be explained
if QCD radiative corrections are properly taken into account. The
exact formulae at order $\alpha_s$ are known since last year, and
only very recently have correct numerical analyses of these formulae
been presented. As the situation is now, the experimental results for
the semileptonic branching ratio and for the charm-counting rate
obtained in high-energy measurements are in perfect agreement with
theory (see Fig.~\ref{fig:BSL}), whereas the results of low-energy
measurements can be explained at the $1\sigma$ level by using a low
renormalization scale.

In order to obtain a detailed understanding of beauty lifetimes, it
is necessary to go to order $1/m_b^3$ in the heavy-quark expansion,
at which the matrix elements of four-quark operators appear. They
describe the physics of spectator effects, i.e.\ contributions in
which a light quark in a beauty hadron is actively involved in the
weak interaction. We have presented a model-independent study of such
contributions, introducing a minimal set of hadronic parameters,
which eventually may be determined using some field-theoretic
approach such as lattice gauge theory. We find that in $B$-meson
decays the coefficients of the colour-octet non-factorizable
operators are much larger than those of the colour-singlet
factorizable operators, and therefore the contributions from the
non-factorizable operators cannot be neglected. The theoretical
prediction for the ratio $\tau(B^-)/\tau(B^0)$ is in agreement with
experiment; however, our present ignorance about the matrix elements
of four-quark operators does not allow us to calculate this ratio
with an accuracy of better than about 20\%. The short $\Lambda_b$
lifetime, on the other hand, remains a potential problem for the
heavy-quark theory. If the current experimental value persists,
either some hadronic matrix elements of four-quark operators must be
significantly larger than naive expectations, or (local)
quark--hadron duality fails in non-leptonic inclusive decays. We
stress that at present the first possibility is not yet ruled out,
although it requires large values of at least some hadronic matrix
elements in the baryon and/or meson sector. In the second case, the
explanation of the puzzle of the $\Lambda_b$ lifetime lies beyond our
present capabilities, as there is no known way to estimate duality
violations in a quantitative way.

\vspace{0.3cm}
{\it Acknowledgements:\/}
Much of the work presented here has been done in a most enjoyable
collaboration with Chris Sachrajda, which is gratefully acknowledged.
I would also like to thank the Organizers of the 1996 Rencontres de
Physique de la Vall\'ee d'Aoste for providing an excellent
environment for a most stimulating conference.

\end{document}